\documentclass[prd,showpcs,amsmath,amssymb,nofootinbib,longbibliography,twocolumn,showpacs,notitlepage]{revtex4-1}
\usepackage{multirow}
\usepackage{epsfig}
\usepackage{amsmath}
\usepackage{bm} 
\usepackage{times}
\usepackage{graphicx}
\usepackage{color}
\usepackage{slashed}
\usepackage{graphicx}
\usepackage{amsmath} 
\usepackage[latin1]{inputenc}
\usepackage{hyperref}
\usepackage{soul}
\usepackage{multirow}
\usepackage{epsfig}
\usepackage{amsmath}
\usepackage{bm} 
\usepackage{times}
\usepackage{graphicx}
\usepackage{color}
\usepackage{slashed}
\usepackage{graphicx}
\usepackage{amsmath}
\usepackage{tikz}
\usetikzlibrary{positioning,shapes}
\usepackage{relsize} 
\usepackage[latin1]{inputenc}

\def\bea{\begin{eqnarray}}
\def\eea{\end{eqnarray}}
\def\bean{\begin{eqnarray*}}
\def\eean{\end{eqnarray*}} 
\def\nn{\nonumber}
\def\beaal{\begin{align}}
\def\eeaal{\end{align}}

\begin{document} 
 
\title{Shedding Light on Dark Sectors with Gravitational Waves}

\author{Nelleke~Bunji\vspace{1mm}}
\affiliation{Department of Chemistry and Physics, Barry University, Miami Shores, Florida 33161, USA\vspace{1mm}}
\author{Bartosz~Fornal}
\affiliation{Department of Chemistry and Physics, Barry University, Miami Shores, Florida 33161, USA\vspace{1mm}}
\author{Kassandra~Garcia\vspace{1mm}}
\affiliation{Department of Chemistry and Physics, Barry University, Miami Shores, Florida 33161, USA\vspace{1mm}}

\date{\today}

\begin{abstract}
\vspace{0mm}The nature of dark matter remains one of the greatest unsolved mysteries in elementary particle physics. It might well be  that the dark matter particle belongs to a dark sector completely secluded or extremely weakly coupled to the visible sector. We demonstrate that gravitational waves arising from first order phase transitions in the early Universe can be used to look for signatures of such dark sector models connected to neutron physics. This introduces a new connection between gravitational wave physics and nuclear physics experiments. Focusing on two particular extensions of the Standard Model with dark U(1) and SU(2) gauge groups constructed to address the neutron lifetime puzzle, we show how those signatures can be searched for in future gravitational wave and astrometry experiments.\vspace{10mm}
 \end{abstract}

\maketitle

\section{Introduction}

Elementary particle physics is a unique area of science in the sense that  it attempts to answer the most fundamental questions about the Universe and its basic constituents. The currently accepted Standard Model works extremely well and precisely describes interactions of the known particles in the visible Universe. It has withstood experimental tests for the last fifty years since its  formulation in the 1970s \cite{Glashow:1961tr,Higgs:1964pj,Englert:1964et,Weinberg:1967tq,Salam:1968rm,Fritzsch:1973pi,Gross:1973id,Politzer:1973fx}, culminating in the discovery of the Higgs particle in 2012 at the Large Hadron Collider \cite{CMS:2012qbp,ATLAS:2012yve}. 

Nevertheless, the discovery of dark matter from  galactic rotation curves in the 1970s \cite{1970ApJ...159..379R} constantly reminds us that  this is not the end of the story. Its existence has been confirmed through various other observations, including  cosmic microwave background \cite{Boomerang:2000efg} and gravitational lensing \cite{Gavazzi:2007vw}. However, the nature of dark matter remains a puzzle and  is certainly one of the most intriguing and crucial questions to answer. We do not even know whether it is an elementary particle or a macroscopic object. The mass of an elementary  particle  could be anywhere between  ${\sim 10^{-31}  \ \rm GeV}$ (fuzzy dark matter \cite{Press:1989id,Hui:2016ltb}) and ${\sim 10^{19} \ \rm  GeV}$ (WIMPzillas \cite{Kolb:1998ki,Meissner:2018cay}), with many well-motivated candidates situated closer to the center of this mass spectrum \cite{Feng:2010gw}. 
If macroscopic, dark matter objects can have masses between  ${\sim 10^{17} \, \rm GeV}$ (dark quark nuggets \cite{Bai:2018dxf})  and  ${\sim 10^{59}  \ \rm GeV}$ (primordial black holes \cite{Carr:2009jm,Bird:2016dcv}).  Thus far, searches at the Large Hadron Collider, direct detection experiments (such as XENONnT \cite{XENON:2015gkh}) and indirect detection (e.g., Fermi Satellite \cite{Fermi-LAT:2017opo}) produced only upper limits on the dark matter couplings to the visible sector.

In light of the null dark matter search results, especially for some of the  best-motivated candidates such as the weakly interacting massive particles (WIMPs) in the $\mathcal{O}(\rm 100 \, \rm GeV)$ mass regime, one needs to use any theoretical and experimental hints currently available. From a theoretical perspective, it is intriguing  that the abundances of dark matter and ordinary matter are of the same order, which suggests that the two sectors might be related  and perhaps share a common origin. Theories of asymmetric dark matter are based on this assertion, and turn this observation into  a prediction for the dark matter mass to be $\mathcal{O}(\rm GeV)$, at the same time explaining the matter-antimatter asymmetry of the Universe  \cite{Nussinov:1985xr,Kaplan:1991ah,Hooper:2004dc,Kaplan:2009ag,Petraki:2013wwa,Zurek:2013wia}. 
This mass range for the dark matter particle has triggered increased interest also on the experimental side, as it was demonstrated that an $\mathcal{O}(1 \ \rm GeV)$ dark matter particle might appear in the final state of the new dark decay channel of the neutron, providing a possible explanation of the neutron lifetime anomaly \cite{Fornal:2018eol} in nuclear physics experiments.

This anomaly arises from the discrepancy between two  qualitatively different types of measurements of the neutron lifetime. 
In the first type of experiments, the bottle method, ultracold neutrons are trapped in a 
container with it interior padded with neutron-reflecting material. The container is then emptied at various storage times and the number of remaining neutrons  is determined using a proportional counter. 
An exponential decay curve is then fitted to those data points, and the neutron lifetime is extracted from the fit. The average of the bottle neutron lifetime experiments  \cite{Serebrov:2004zf,Pichlmaier:2010zz,Steyerl:2012zz,Arzumanov:2015tea,Serebrov:2017bzo,Pattie:2017vsj,Ezhov:2014tna,UCNt:2021pcg}  is  $\tau_n^{\rm bottle} = 878.4 \pm 0.5 \ \rm s$. 
In the second type of measurements, the beam method, a  beam of cold neutrons passes through
a quasi-Penning trap, which collects and counts the protons from neutron decays, enabling the determination of the neutron decay rate involving
protons in the final state. Estimating  the number of neutrons in the beam, the neutron lifetime is  calculated by dividing this number by the rate of decay to protons. The average of the beam neutron lifetime measurements  \cite{Byrne:1996zz,Nico:2004ie,Yue:2013qrc,Hirota:2020mrd}  is $\tau_n^{\rm beam} = 888.0 \pm 2.0 \ {\rm s}$, which is four standard deviations away from the  bottle result.

 Although this mismatch may be due to some unknown systematic errors, it was  demonstrated in \cite{Fornal:2018eol}
that a neutron decay channel into  dark particles can account for this. In particular, if the branching fraction for neutron beta decay is $99\%$, whereas the remaining $1\%$ of decays involves particles from a dark sector with no protons in the final state, the two experimental results are reconciled.
This proposal for the existence of a neutron dark decay channel  was followed up by a plethora of theoretical work \cite{Baym:2018ljz,McKeen:2018xwc,Motta:2018rxp,Barducci:2018rlx,Grinstein:2018ptl,Cline:2018ami,Karananas:2018goc,Bringmann:2018sbs,Berezhiani:2018eds,Jin:2018moh,Keung:2019wpw,Elahi:2020urr,Fornal:2020bzz,McKeen:2020zni,Alonso-Alvarez:2021oaj}, as well as experimental efforts and proposals \cite{Tang:2018eln,Sun:2018yaw,Pfutzner:2018ieu,Klopf:2019afh,Ayyad:2019kna,Ayyad:2022zqw,Lopez-Saavedra:2022vxh,Tang,2021APS..DNP.QJ008H}; for a review see \cite{Fornal:2023wji}. In this paper we complement this literature by analyzing the possible gravitational wave signatures from UV complete theories  constructed for the neutron dark decay channel,  focusing on the models constructed in \cite{Cline:2018ami,Elahi:2020urr}.

Recently a novel and very promising avenue of  probing particle physics models  has emerged with the first confirmed detection of gravitational waves by the Laser Interferometer Gravitational Wave
Observatory (LIGO) \cite{LIGOScientific:2016aoc}. Although this signal and some $\mathcal{O}(100)$ subsequent events arose from mergers of black holes and/or neutron stars, gravitational waves could have been produced in the early Universe as well. 
 Among  the most spectacular  hypothetical processes occurring shortly after the Big Bang and leading to a potentially measurable stochastic gravitational wave background today are: first order phase transitions \cite{Kosowsky:1991ua}, cosmic strings \cite{Vachaspati:1984gt,Sakellariadou:1990ne}, domain walls \cite{Hiramatsu:2010yz}, and  inflation  \cite{Turner:1996ck}.

 First order phase transitions are especially interesting, since they strongly depend on the particle physics details. They are triggered when the effective potential  develops a  deeper minimum (true vacuum) at a nonzero field value separated from the high temperature minimum (false vacuum) by a potential barrier. When a given point in the Universe transitions from the false vacuum to the true one, this corresponds to a nucleation of an expanding bubble of true vacuum. Such a process can happen in multiple points in space, eventually causing the entire Universe to transition to the true vacuum state. During the bubble nucleation and expansion, gravitational waves are emitted from sound waves in the primordial plasma, bubble collisions, and turbulence. 
 There is vast literature on the subject  analyzing a plethora of  particle physics models, including new electroweak-scale physics
\cite{Grojean:2006bp,Vaskonen:2016yiu,Dorsch:2016nrg,Bernon:2017jgv,Baldes:2018nel,Chala:2018ari,Alves:2018jsw,Han:2020ekm}, dark sectors \cite{Schwaller:2015tja,Baldes:2017rcu,Breitbach:2018ddu,Croon:2018erz,Hall:2019ank, Fornal:2022qim,Kierkla:2022odc},  axions \cite{Dev:2019njv,VonHarling:2019rgb,DelleRose:2019pgi,ZambujalFerreira:2021cte},
unification \cite{Croon:2018kqn,Huang:2020bbe,Okada:2020vvb,Fornal:2023hri}, conformal invariance \cite{Ellis:2020nnr,Kawana:2022fum}, supersymmetry \cite{Craig:2020jfv,Fornal:2021ovz}, left-right symmetry \cite{Brdar:2019fur,Graf:2021xku},
neutrino mass models \cite{Brdar:2018num,Okada:2018xdh,DiBari:2021dri,Zhou:2022mlz}, baryon and lepton  number violation \cite{Hasegawa:2019amx,Fornal:2020esl,Bosch:2023spa}), flavor physics \cite{Greljo:2019xan,Fornal:2020ngq}, and  leptogenesis \cite{Dasgupta:2022isg}. For a review of the subject see, e.g.,  \cite{Caldwell:2022qsj} and for the LIGO observing run O3 constraints on particle physics models see \cite{LIGO_FOPT}.

The reach of gravitational wave observations will improve considerably with future detectors, such as the 
Laser Interferometer Space Antenna  (LISA) \cite{Audley:2017drz},  Cosmic Explorer (CE) \cite{Reitze:2019iox}, Big Bang Observer (BBO) \cite{Crowder:2005nr},  Einstein Telescope (ET) \cite{Punturo:2010zz}, and DECIGO \cite{Kawamura:2011zz}. 
Simultaneously to those gravitational wave measurements using interferometers, there are several existing and upcoming observational efforts to detect gravitational waves through their effect on pulsar timing arrays,  sensitive to much lower frequencies. This includes: NANOGrav \cite{NANOGRAV:2018hou}, PPTA \cite{2013PASA...30...17M}, EPTA \cite{2010CQGra..27h4014F}, IPTA \cite{2010CQGra..27h4013H}, and SKA \cite{Weltman:2018zrl}. It is worth noting that quite recently NANOGrav has  detected a stochastic gravitational wave signal in the $\sim 10^{-8} \ \rm Hz$ frequency region \cite{NANOGrav:2023hvm}.

The aim of this paper is to demonstrate that gravitational waves experiments grant access to yet unexplored parameter space of models relevant for nuclear physics. For concreteness, we focus on two models constructed for the  neutron dark decay and describe them in Sections \ref{model12} and  \ref{model22}, including  constraints  from cosmological and astrophysical observations, as well as direct and indirect detection experiments.  Section \ref{pt} analyzes the first order phase transition in each model. Finally, in Section \ref{gws} we derive the expected gravitational wave signatures and comment on their relation to the recent NANOGrav signal. Our findings are summarized in Section \ref{sum}.

\section{Model 1} \label{model12}

The model we first consider is based on the  symmetry \cite{Cline:2018ami} 
\bea\label{symmetry}
{\rm SU}(3)_c \times {\rm SU}(2)_L \times {\rm U}(1)_Y \times {\rm U}(1)_D \ ,
\eea
with the dark ${\rm U(1)}_D$ gauge group spontaneously   broken at the energy scale $\mathcal{O}(60\, \rm MeV)$. The neutron dark decay channel proposed in \cite{Cline:2018ami}  is $n \to \chi\,A'$, where $\chi$ is a dark fermion  and $A'$ is a dark photon. 

In this section  we complement this analysis  by considering, within the framework of this model,  the neutron dark decay channel $n \to \chi \,\phi$, where $\phi$ is the scalar responsible for  ${\rm U}(1)_D$ symmetry breaking. As will be shown in Section \ref{gws},  this scenario  leads to a measurable  gravitational wave signal.

\subsection{Particle content and Lagrangian}

The theory extends the Standard Model by introducing the following new fields:
\begin{itemize}
\item[$\bullet$]  Complex scalar $\phi = (1,1,0,1)$ with baryon number $B_\phi = 0$, responsible for the breaking of ${\rm U(1)}_D$;
\item[$\bullet$] Dirac fermion  $ \chi =  (1,1,0,1)$ carrying $B_\chi=1$,  a dark  particle  and product of neutron dark decay;
\item[$\bullet$]  Dark photon  $A'$ from ${\rm U(1)}_D$ breaking, which alleviates neutron star constraints;
\item[$\bullet$]  Two complex scalars triplets:  $\Phi_1 = (3,1,\tfrac13,-1)$ and  $\Phi_2 = (3,1,\tfrac13,0)$  with $B_\Phi = -2/3$, providing the interactions  for the neutron dark decay to be triggered. 
\end{itemize}

The beyond-Standard-Model  Lagrangian terms relevant for neutron dark decay and generating masses for $\chi$, $A'$, $\phi$ are:
\bea
\mathcal{L}  &\supset& |D_\mu \phi|^2 +  \lambda\left[|\phi|^2 - \big(\tfrac{v_D}{\sqrt2}\big)^2\right]^2+  \bar{\chi}\,\big(i\slashed D-m_\chi\big) \,\chi   \nn\\
&+&\lambda_1 \bar{d}^{\,i} P_L \chi \Phi_{1i} + \lambda_2 \epsilon^{ijk} \bar{u}^c_i P_R d_j \Phi_{2k} \nn\\
&+& \mu \,\Phi_{1i} \Phi_2^{*i} \phi  - \tfrac14 F'_{\mu\nu} F'^{\mu\nu} - \tfrac\delta 2 F_{\mu\nu} F'^{\mu\nu} \ ,
\eea
where the covariant derivative is $D_\mu = \partial_\mu - i g_D A'_\mu$ and $i,j,k$ are color indices. A nonzero $\delta$  parameter allows for a kinetic mixing between $A'$ and the photon, leading to the decay channel $A' \to e^+e^-$. 

Upon spontaneous breaking of the ${\rm U(1)}_D$ gauge symmetry when $\phi$ develops the vacuum expectation value $\langle \phi \rangle = v_D/\sqrt2$, the  dark photon  and scalar $\phi$ acquire the  masses:
\bea\label{eqq3}
\begin{gathered}
m_{A'} = \frac{g_Dv_D}{\sqrt2} \ ,\\
\ \ \ \ m_\phi = \sqrt{2\lambda} \,v_D \ .
\end{gathered}
\eea
We assume that the masses of the scalars $\Phi_1$ and $\Phi_2$ are generated at a higher scale, and are much larger than the masses of  $\chi$, $A'$, $\phi$, so they can be integrated out. 

\subsection{Low-energy effective theory}

\begin{figure}[t!]
\includegraphics[trim={0cm 2.5cm 0cm 2.5cm},clip,width=6cm]{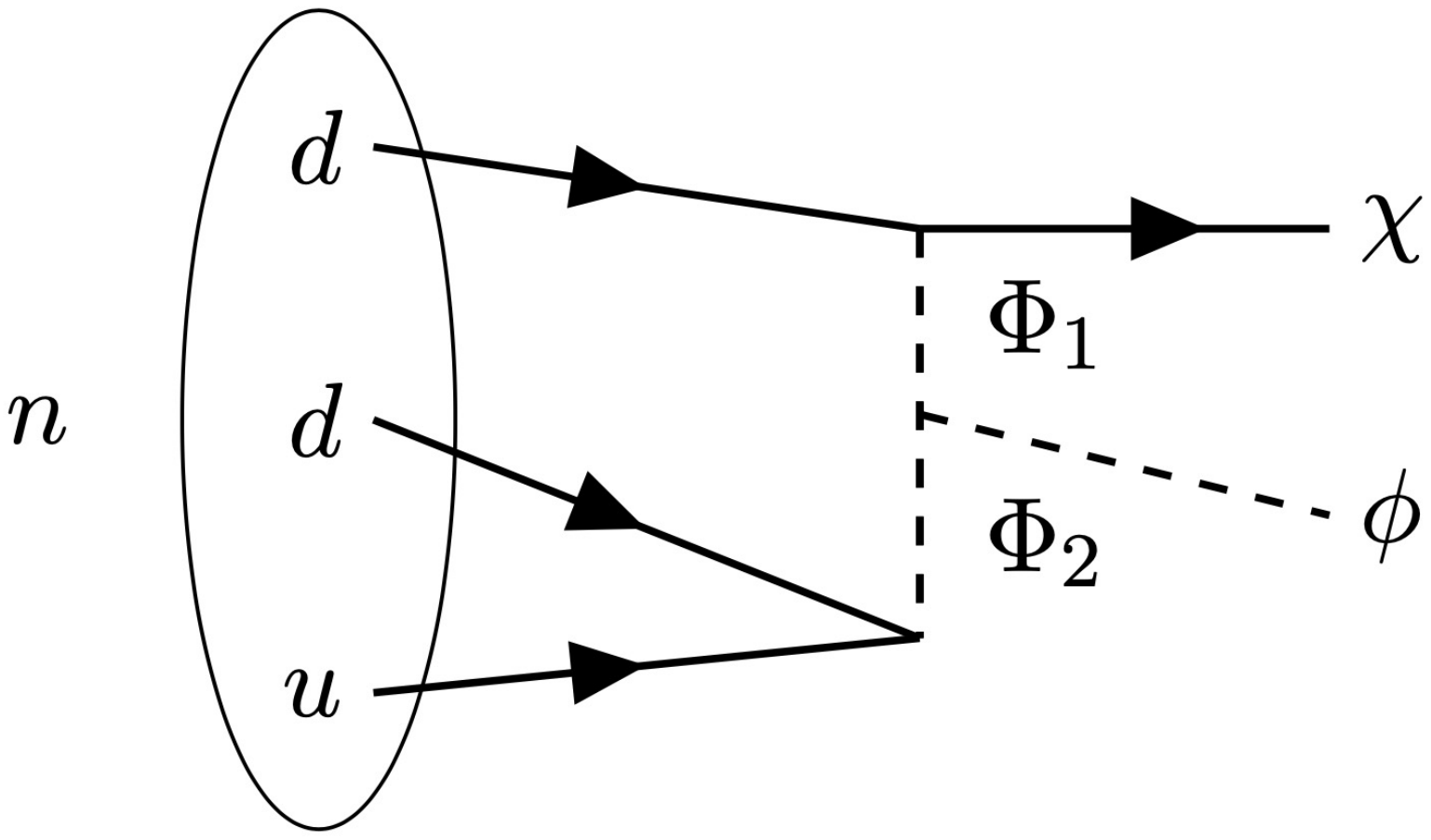} \vspace{-2mm}
\caption{Possible neutron dark decay channel $n \to \chi\,\phi$ in the dark ${\rm U}(1)_D$ model \cite{Cline:2018ami} and in the dark ${\rm SU}(2)_D$ model \cite{Elahi:2020urr}.}\label{fig:1}
\end{figure}

Below the mass scale of $\Phi_1$ and $\Phi_2$, an effective coupling of the dark particle $\chi$ to the neutron and $\phi$ is generated,
\bea
\mathcal{L}_{n\chi\phi} = \frac{\varepsilon \sqrt2}{v_D} \, \bar{n} P_L \chi \,\phi\ ,
\eea
where
\bea
\varepsilon = \frac{\beta\, \mu\, v_D \lambda_1 \lambda_2}{m_{\Phi_1}^2 m_{\Phi_2}^2} \ ,
\eea
with  $\beta$ being the matching coefficient determined from a lattice calculation, $\beta =  0.0144(3)(21) \ \rm GeV^3$ \cite{Aoki:2017puj}.
Therefore, after ${\rm U(1)}_D$  breaking, the nonstandard contribution to the effective Lagrangian at the nuclear level is
 \bea\label{effL}
\mathcal{L}_{\rm eff} &=&
  \bar{\chi}\,\big(i\slashed D-m_\chi\big) \,\chi + \varepsilon \left(\bar{n}\,\chi + \bar{\chi}\,n\right) -  \frac12 m^2_{A'} {A'}^\mu \!A'_{\mu} \nn\\
&-& \frac14 F'_{\mu\nu} F'^{\mu\nu} - \frac\delta 2 F_{\mu\nu} F'^{\mu\nu}  + \frac{\varepsilon \sqrt2}{v_D}  \left(\bar{n}\,\chi + \bar{\chi}\,n\right) \phi \ . \ \ \ \ \ \ 
\eea
Depending on the details of the spectrum of the theory, those interactions can lead to  neutron dark decay. The focus of \cite{Cline:2018ami} was on $n \to \chi \, A' $, which would yield a rate of
\bea\label{drateold}
\Gamma(n\to \chi A') = \frac{g_D^2 \varepsilon^2}{8\pi} \frac{(m_n \!-\! m_\chi)}{m^2_{A'}} \left(1-\frac{m^2_{A'}}{(m_n\!-\!m_\chi)^2}\right)^{\frac32} \!\!\!\! . \ \ \ \ \ 
\eea
However, the other possibility within this model, not analyzed in \cite{Cline:2018ami}, is the decay channel  $n \to \chi \, \phi$ (see, Figure \ref{fig:1}). We find that the corresponding neutron dark decay rate is
\bea\label{drate}
\Gamma(n\to \chi\,\phi) &=& \frac{\varepsilon^2}{8\pi v_D^2m_n} \sqrt{\left( m_n-{m_\chi}\right)^2 - {m_\phi^2}} \nn\\
&\times&\sqrt{\left( m_n+{m_\chi}\right)^2 - {m_\phi^2}} \ .
\eea

As discussed below, there exists a range of parameter values for which this decay rate corresponds to a neutron decay branching fraction of 1\%, needed to explain the neutron lifetime discrepancy, while remaining consistent with all experimental and observational constraints.

\subsection{Phenomenology}\label{constr}

Given the complexity of the relations between parameters in the model, we  focus on particular benchmark scenarios with fixed masses for $\chi$ and  $\phi$. The constraints arise from  various cosmological and astrophysical observations, but  are not very sensitive to the precise choices of the particle masses. \vspace{4mm}

\noindent 
$(a)$ Neutron dark decay \vspace{2mm}

To prevent neutrons in stable nuclei from undergoing dark decays, one requires \cite{Pfutzner:2018ieu}
\bea
937.993 \, {\rm MeV} < m_\chi + m_{\phi} < 939.565 \, {\rm MeV} \ ,
\eea
where the lower bound is slightly higher than in the  original proposal \cite{Fornal:2018eol} due to taking into account a
rapid disintegration of excited $^8{\rm Be}$  into two $\alpha$ particles. The dark particle $\chi$ is assumed to be Dirac to avoid bounds from dinucleon decay searches. 
One of the benchmark points $(\rm A)$ we consider is
\bea\label{benchaaa}
m_\chi = 938 \ {\rm MeV}  ,\  \ m_\phi = 0.85 \ {\rm MeV}  , \ \ m_{A'} = 10 \ \rm MeV. \ \ \ \ 
\eea
In order to resolve the neutron lifetime discrepancy through a $1\%$ branching fraction for the dark decay \cite{Fornal:2018eol}, one needs  $\Gamma(n\to \chi \,\phi) \approx 7 \times 10^{-30} \ \rm GeV$. For benchmark $(\rm A)$   this is equivalent to having
\bea\label{eps}
\varepsilon \approx 1.6 \times 10^{-11} \ \rm MeV \ .
\eea

\noindent 
$(b)$ Neutron stars \vspace{2mm}

Constraints from observed neutron star masses \cite{Baym:2018ljz,McKeen:2018xwc,Motta:2018rxp} on the models originally proposed in \cite{Fornal:2018eol} are alleviated by self-interactions in the dark sector  mediated by the dark photon. In particular, the resulting repulsive interaction between $\chi$ and the neutrons make the neutron star equation of state stiffer. To account for  masses of $2M_\odot$, i.e., the largest neutron star masses observed, it is sufficient to require \cite{Cline:2018ami}
\bea\label{nns}
\frac{v_D}{\sqrt2} = \frac{m_{A'}}{g_D} \lesssim (45 - 60) \ \rm MeV \ ,
\eea
depending on the assumed nuclear star equation of state. In our analysis we fulfill Eq.\,(\ref{nns}) by setting $v_D = 60 \ \rm MeV$.
\vspace{4mm}

\noindent 
$(c)$ Cosmology \vspace{2mm}

Further constraints arise from Big Bang nucleosynthesis, measurements of the cosmic microwave background, and supernova observations. Those bounds strongly disfavor a dark photon $A'$ with mass less than $2\,m_e$. As shown in  \cite{Cline:2018ami}, for $m_{A'} = 1.35 \ \rm MeV$ the allowed range for the parameter $\delta$ is limited to  $ 2 \times 10^{-11} < \delta < 2 \times 10^{-9}$,  while the dark gauge coupling $g_D > 0.07$. Those bounds are fairly independent of the  value of $m_{A'}$, as long as  $m_{A'} > 2 \,m_e$.

In the scenario we consider, i.e., $n \to \chi\,\phi$, the dark photon mass does not need to be small. Increasing $m_{A'}$ loosens the cosmological and astrophysical bounds on the model. In addition, we are not considering the case when the dark photon  or the dark scalar are the dark matter, which eliminates direct and indirect dark matter detection bounds.

\vspace{1mm}
Furthermore,  the case of a heavier dark photon is preferred for the gravitational wave signals. 
Based on Eq.\,(\ref{eqq3}), the ratio of the dark photon mass and dark scalar mass is
\bea
\frac{m_{A'}}{m_\phi} = \frac{g_D}{2\sqrt{\lambda}} \ ,
\eea
while gravitational waves from first order phase transitions, as discussed will be in Section \ref{gws}, provide strongest signals when $g_D/(2\sqrt{\lambda}) > \mathcal{O}(1)$, which corresponds precisely to the case of a  heavier dark photon.

We also note that in the gravitational wave analysis of the model we  remain general and go beyond the parameter space allowing for a neutron dark decay channel, e.g., we consider also $\phi$ and $A'$ masses in the multi-MeV range.

\section{Model 2} \label{model22}

We now turn to  the model based on the gauge group \cite{Elahi:2020urr} 
\bea\label{symmetry2}
{\rm SU}(3)_c \times {\rm SU}(2)_L \times {\rm U}(1)_Y \times {\rm SU}(2)_D \ ,
\eea
where  ${\rm SU(2)}_D$ is broken at energies $\mathcal{O}(60\, \rm MeV)$. There are three scenarios for the 
 neutron dark decay proposed in \cite{Elahi:2020urr}, but we focus on  $n \to \chi\,\phi$, where $\chi$ is a dark fermion  and $\phi$ is the scalar responsible for ${\rm SU}(2)_D$ breaking. In 
 Section \ref{gws} we will demonstrate that this  scenario can be probed with gravitational waves and discuss how this case differs from the Abelian model described in Section \ref{model12}.

In its structure, Model 2 is very similar to Model 1, but with three differences:
\begin{itemize}
\item[$\bullet$]  Complex scalar $\Phi = (1,1,0,2)$ responsible for the breaking of ${\rm SU(2)}_D$ is a dark doublet,
\bea
\Phi =\begin{pmatrix}
           \tfrac{1}{\sqrt2}(G_1 + i G_2) \\
           \tfrac{\phi + v_D}{\sqrt2} + i G_3 
         \end{pmatrix};
\eea
\item[$\bullet$]  Upon symmetry breaking, instead of one dark photon there are three dark ${W}^{\prime a}$ gauge bosons ($a=1,2,3$);
\item[$\bullet$]  Complex scalar triplet $\Phi_1$ is now an ${\rm SU}(2)_D$ doublet:  $\Phi_1 = (3,1,\tfrac13,2)$.
\end{itemize}

The Lagrangian takes the form,
\bea\label{LagnonA}
\mathcal{L}  &\supset& |D_\mu \Phi|^2 +   \lambda\left[|\Phi|^2 - \big(\tfrac{v_D}{\sqrt2}\big)^2\right]^2+  \bar{\chi}\,\big(i\slashed D-m_\chi\big) \,\chi   \nn\\
&+&\Big[\lambda_1 \bar{d}^{\,i} P_L \chi \Phi_{1i} + \lambda_2 \epsilon^{ijk} \bar{u}^c_i P_R d_j \Phi_{2k}   \nn\\
&+&  \mu \,\Phi_{1i} \Phi_2^{*i} \Phi + {\rm h.c.}\Big]- \tfrac14 {W}^{\prime a}_{\!\!\mu\nu} W^{\prime a,\mu\nu} \nn\\
&+& \Big[c_1 {\rm Tr}(\Phi^\dagger \tau^a \Phi  \,{W}^{\prime a}_{\!\!\mu\nu}{F}^{\mu\nu})+ c_2 {\rm Tr}(\Phi^\dagger \tau^a \Phi  \,{W}^{\prime a}_{\!\!\mu\nu}\widetilde{F}^{\mu\nu})\nn\\
&+&  c_3 {\rm Tr}(\Phi^\dagger \tau^a \Phi  \,{\widetilde{W}}^{\prime a}_{\!\!\mu\nu}{F}^{\mu\nu}) + {\rm h.c.}\Big] \ , \ \ \ \ \ \ 
\eea
where the covariant derivative is $D_\mu = \partial_\mu - i g_D \tau^a W^a_\mu$,\break ${W}^{\prime a}_{\!\!\mu\nu}$ is the ${\rm SU(2)}_D$ field strength tensor, and its dual is given by ${\widetilde{W}}^{\prime a}_{\!\!\mu\nu}= \epsilon_{\mu\nu\alpha\beta}{W}^{\prime a,\alpha\beta}$.

Upon spontaneous breaking of ${\rm SU(2)}_D$ when $\Phi$ develops its vacuum expectation value, 
\bea
\langle \Phi \rangle =\tfrac{1}{\sqrt2}\begin{pmatrix}
          0\\
            v_D  
         \end{pmatrix},
\eea
the  dark $W^{\prime a}$ gauge bosons  and the radial component of the scalar $\Phi$ acquire the following masses,
\bea\label{eqq4}
\begin{gathered}
m_{W'} = \frac{g_D v_D}{\sqrt2} \ ,\\
\ \ \ \ \ m_\phi = \sqrt{2\lambda} \,v_D \ .
\end{gathered}
\eea
Assuming again that the masses of the scalars $\Phi_1$ and $\Phi_2$ are generated at a higher scale than $v_D$, and upon integrating them out, one arrives at a similar low-energy effective theory as for Model 1 in Eq.\,(\ref{effL}), but with the dark photon terms replaced by those containing the $W^{\prime a}$ gauge boson fields.

Depending on the masses of the particles,  the possible dark decay channels for the neutron are $n\to \chi\,W'$ and $n\to\chi\,\phi$. For similar reasons as before, i.e., to have a scenario with 
measurable gravitational wave signals, we assume 
\bea\label{masses2}
\frac{m_{W'}}{m_\phi} = \frac{g_D}{2\sqrt{\lambda}}  > \mathcal{O}(1)\ .
\eea
Consequentially, we focus again on  the $n\to\chi\,\phi$ dark decay channel (see, Figure \ref{fig:1}), since it is realized when the condition in Eq.\,(\ref{masses2}) is fulfilled. The formula for the decay rate is identical to the one in Eq.\,(\ref{drate}), and there is again a range of possible parameter values for which this decay rate yields a neutron dark decay branching fraction of 1\%.

We consider the benchmark point $(\rm A)'$, similar to $(\rm A)$ for Model 1, but with  the dark photon replaced by $W'$,
\bea\label{20}
m_\chi = 938 \ {\rm MeV}  ,\  \, m_\phi = 0.85 \ {\rm MeV}  , \  \, m_{W'} = 8 \ \rm MeV  , \ \ \ \
\eea
which results
 in the same value of $\varepsilon$ as the one  in Eq.\,(\ref{eps}). 
Neutron star constraints  translate to
\bea
\frac{v_D}{\sqrt2} = \frac{m_{W'}}{g_D} \lesssim (45 - 60) \ \rm MeV \ ,
\eea
and  we again  take $v_D = 60 \ \rm MeV$.

It was argued in \cite{Elahi:2020urr}, in agreement  with \cite{Cline:2018ami}, that $\chi$  cannot be the sole component of dark matter. At the same time, it was demonstrated that  the dark matter can be made up of a combination  of $\phi$ and $W^{\prime a}$, or just the $W^{\prime a}$, depending on the region of parameter space. 
Those dark matter particles would be produced via the freeze-in mechanism \cite{Elahi:2020urr}.

Constraints arising from cosmology, astrophysics (other than neutron stars) and dark matter direct detection are all less severe than the requirement of  the correct dark matter relic density. In particular, the Big Bang nucleosynthesis bounds are mild as long as $m_{W'} > 2 m_e$, the cross section for indirect dark matter detection is extremely small and does not provide any noteworthy bound, and  the direct dark matter detection constraints are even less constraining. 

The  cosmological and astrophysical bounds on the model are not very sensitive  to $m_\phi$, leading to similar exclusion regions  for our benchmark point as in scenario ${\bold{C}}$ in \cite{Elahi:2020urr}, i.e., for the dark gauge coupling $g_D > 0.05$ and the coefficient for the first trace term in Eq.\,(\ref{LagnonA}) $c_1  < (10^{-14},10^{-9})\ {\rm GeV}^{-2}$ as $g_D$ changes in the range $(0.05, 1)$.

\section{First order phase transition }
\label{pt}

We now investigate the possibility of  a first order phase transition being  triggered by ${\rm U}(1)_D$ breaking in Model 1 and  ${\rm SU}(2)_D$ breaking in Model 2 around  the scale $\mathcal{O}(60 \,\rm MeV)$.
We first calculate the  effective potential for both models, and then determine the effective action for the bounce solution. This is then used to  find the phase transition parameters $\alpha$, $\tilde\beta$, and $T_*$, ultimately leading to a prediction for the gravitational wave signal, as  derived in Section \ref{gws}.

\subsection{Effective potential}

The effective potential,  consists of three parts: tree-level  $V_{\rm tree}(\varphi)$, one-loop Coleman-Weinberg correction $V_{\rm loop}(\varphi)$, and the finite temperature contribution $V_{\rm temp}(\varphi,T)$, where $\varphi$ is the background field, 
\bea
V_{\rm eff} (\varphi, T) = V_{\rm tree}(\varphi) + V_{\rm loop}(\varphi) + V_{\rm temp}(\varphi,T)\ . \ \ \ 
\eea
Plugging into the formula for the tree-level potential  the value of $m_\phi$ obtained by minimizing the potential, one arrives at
\bea\label{sc2newnew}
V_{\rm tree}(\varphi) = -\frac12 \lambda \,v^2_D \varphi^2 + \frac14 \lambda \,\varphi^4\ .
\eea
To determine the one-loop contribution, we implement the cut-off regularization scheme and choose the minimum of the potential at zero temperature  and the mass of $\phi$ to be equal to their tree-level values. This leads to \cite{Anderson:1991zb}
\bea
V_{\rm loop}(\varphi) &=& \sum_{i}\frac{n_i}{(8\pi)^2}\bigg\{m_i^4(\varphi) \left[\log \left(\frac{m_i^2(\varphi)}{m_i^2(v_D)} \right)-\frac32\right]\nn\\
&&+ \ 2 \,m_i^2(\varphi) \,m_i^2(v_D) \bigg\} \ ,
\eea
where the sum is over  all particles charged under ${\rm U}(1)_D$   ($ {\rm SU}(2)_D$), including Goldstone bosons, $n_i$ is the number of degrees of freedom for a given particle species, and $m_i$ are the  background field-dependent masses (with the substitution $m_{\chi}(v_D) \to m_{\phi}(v_D)$ for the Goldstones).

We find that the background field-dependent masses for the ${\rm U}(1)_D$ charged particles in Model 1, i.e., the dark photon $A'$, the scalar $\phi$, and the Goldstone boson $\chi$, are
\bea
m_{A'}(\varphi) &=& g_D \varphi\ , \ \ \ \ \ m_{\phi}(\varphi) = \sqrt{\lambda(3\varphi^2 - v_D^2)}\ ,\nn\\ 
m_{\chi}(\varphi) &=&\sqrt{\lambda(\varphi^2 - v_D^2)}\ ,
\eea
with $n_{A'} = 3$ and $n_\phi = n_\chi = 1$.

 In the case of Model 2, the ${\rm SU}(2)_D$ charged particles, i.e., the dark gauge bosons $W^{\prime a}$, the scalar $\phi$, and the Goldstone bosons $\chi^a$, have 
 the same background field-dependent masses and numbers of degrees of freedom as the dark photon, scalar, and Goldstone boson in  Model 1, respectively. The difference, however,  is that there are three $W^{\prime a}$ gauge bosons and three Goldstone bosons in the summation.

The temperature dependent part of the effective potential is given by \cite{Quiros:2007zz}
\bea
&&\hspace{-8mm}V_{\rm temp}(\varphi,T)  \nn\\
&=&\,\frac{T^4}{2\pi^2} \sum_i n_i \int_0^\infty dx\,x^2 \log\left[1\mp e^{-\sqrt{x^2+\tfrac{m_i^2(\varphi)}{T^2}}}\right]\nn\\
&+&\,\frac{T}{12\pi} \sum_k n'_k \left\{m_k^3(\varphi)-\left[m_k^2(\varphi) + \Pi_k(T)\right]^{3/2}\right\} . \ \ \ \ 
\eea
The first part of this expression (sum includes all particles) corresponds to one-loop diagrams, whereas the second part (sum includes only bosons) comes from Daisy diagrams, with only longitudinal degrees of freedom for vector bosons  involved. The thermal masses $\Pi_k(T)$ are calculated using the formalism presented, e.g., in \cite{Comelli:1996vm}. Assuming $\lambda \ll 1$, which is fulfilled for the parameter space considered, we obtain the results below.

\begin{figure}[t!]
\includegraphics[trim={1.0cm 0.6cm 1cm 0cm},clip,width=8cm]{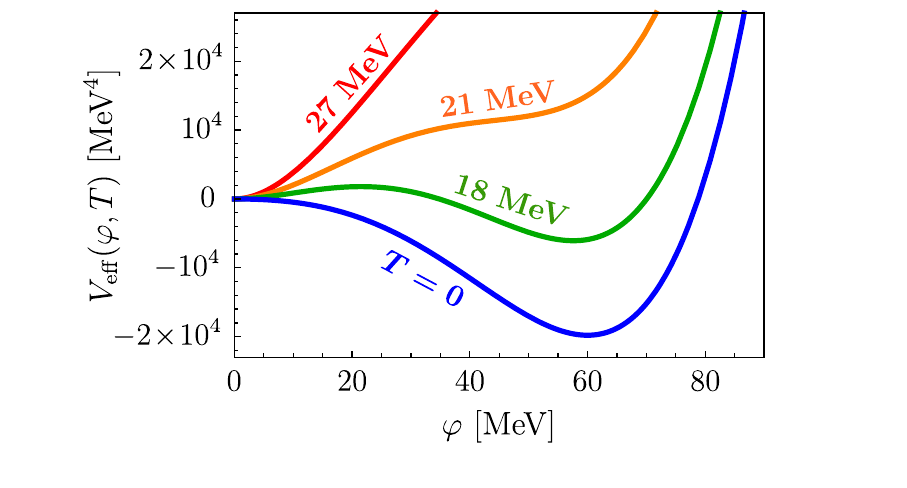} \vspace{-3mm}
\caption{Effective potential for the ${\rm U(1)}_D$  model assuming $g_D=0.8$, $v_D = 60 \ \rm MeV$, and $\lambda = 0.01$,  plotted for several  temperatures.}\label{fig:2}
\end{figure}

The thermal masses for the dark photon $A'$, scalar $\phi$, and Goldstone boson in Model 1 are,
\bea
\Pi_{A'}(T) &=& \frac13 g_D^2 T^2, \nn\\
\Pi_{\phi}(T) &=& \Pi_{\chi}(T)   = \frac14 g_D^2 T^2 . \ \ \ \ \ \ 
\eea
The values for $n_i$ are the same as in the zero temperature  calculation, whereas $n'_{A'} = 1$ and  $n'_\phi = n'_\chi = 1$.

In the  ${\rm SU}(2)_D$ scenario of Model 2,  the thermal masses are,
\bea
\Pi_{W'}(T) &=& \frac56 g_D^2 T^2\ ,\nn\\
 \Pi_{\phi}(T) &=& \Pi_{\chi}(T)   = \frac34 g_D^2 T^2 . \ \ \ \ \ \ 
\eea
In this case $n_{W'} = 3$, $n'_{W'} = 1$, $n_\phi = n_\chi = n'_\phi = n'_\chi =  1$, and there are again three $W^{\prime a}$ and three $\chi^a$ fields.\vspace{2mm}

The effective potential for Model 1, assuming  $g_D=0.8$, $v_D = 60 \ \rm MeV$,  $\lambda = 0.01$, and several values of the temperature is plotted in Figure \ref{fig:2}. As the temperature drops, a new true vacuum at $\varphi \ne 0$ appears with lower energy density than the $\varphi = 0$ false vacuum. The two vacua are separated by a potential barrier, which may result in a first order phase transition triggering bubble nucleation.

\subsection{Bubble nucleation}

The transition from the false vacuum to the true vacuum is initiated when the  temperature drops below the nucleation temperature $T_*$. When this happens, bubbles of true vacuum are nucleated  in various  parts of the Universe. The nucleation temperature  is calculated by comparing the bubble nucleation rate  $\Gamma(T)$ with the Hubble expansion rate,  since only when the two are comparable the nucleation  will be efficient enough to continue. We are therefore looking for a solution to:
\bea\label{gamma}
\Gamma(T_*) \sim H(T_*)^4 \ ,
\eea
with the nucleation  rate \cite{LINDE1983421}
\bea\label{PT_Gamma}
\Gamma(T) \approx \left(\frac{\mathcal{S}(T)}{2\pi T}\right)^{\frac32}T^4 \exp\left({-\frac{\mathcal{S}(T)}{T}}\right) \ .
\eea
Here $\mathcal{S}(T)$ is  the Euclidean action,
\bea
\mathcal{S}(T)= 4\pi\int dr\,r^2 \left[\frac12\left(\frac{d\varphi_b}{dr}\right)^2+V_{\rm eff}(\varphi_b, T)\right] , \ \ \ \ \ 
\eea
where $\varphi_b(r)$  is the bounce solution determining the profile of the expanding bubble, i.e., solution of the equation,
\bea
\frac{d^2 \varphi}{dr^2}+\frac{2}{r}\frac{d\varphi}{dr}- \frac{d V_{\rm eff}(\varphi,T)}{d\varphi}= 0 \ ,
\eea
subject to the  boundary conditions:
\bea
\frac{d\varphi}{dr}\Big|_{r=0} = 0 \ , \ \ \ \ \ \varphi(\infty) = \varphi_{\rm false} \ .
\eea
Using Eqs.\,(\ref{gamma}) and (\ref{PT_Gamma}), the nucleation temperature $T_*$ is then determined by solving,
\bea\label{nucl_temp}
\frac{\mathcal{S}(T_*)}{T_*}   \approx  4\log\!\left[\frac{M_{Pl}}{T_*}\right]  - 2\log\left[\frac{4\pi^3g_*}{45}\!\left(\frac{2\pi \,T_*}{\mathcal{S}(T_*)}\right)^{\!\!\frac34}\right], \ \ \ \ \ \ 
\eea
where $M_{Pl}$ is the Planck mass.

\subsection{Phase transition parameters}
After computing the nucleation temperature, the parameter $\alpha$ describing the  phase transition strength is determined  from 
\bea\label{alphaA}
\alpha = \frac{\rho_{\rm vac}(T)}{\rho_{\rm rad}(T)}\bigg|_{T=T_*}\ ,
\eea
i.e., the offset between the two vacua's energy densities
\bea
\rho_{\rm vac}(T) &=& V_{\rm eff}(\varphi_{\rm false},T) -  V_{\rm eff}(\varphi_{\rm true},T)\nn\\
&-&T \frac{\partial}{\partial T} {\Big[ V_{\rm eff}(\varphi_{\rm false},T) -  V_{\rm eff}(\varphi_{\rm true},T)\Big]} \ \ \ \ \ \ 
\eea
divided by the energy density of radiation
\bea
\rho_{\rm rad}(T) = \tfrac{\pi^2}{30} g_* T^4 \ .
\eea
The parameter $\tilde\beta$, determining the inverse of the  duration of the phase transition, can be calculated via
\bea\label{beta}
\tilde{\beta} = T_* \frac{d}{dT} \!\left(\frac{\mathcal{S}(T)}{T}\right)\bigg|_{T=T_*} \ .
\eea

\section{Gravitational wave signal}\label{gws}

When bubbles of true vacuum are nucleated and violently expand, the combined effect of collisions between their walls, sound shock waves in the plasma, and magnetohydrodynamic turbulence
gives rise to gravitational waves, which propagate through the Universe and would reach us today in the form of a stochastic gravitational wave background. 

The shape of its spectrum has been successfully modeled through numerical simulations. It depends on 4 parameters: $v_w$ -- speed of the bubble wall which we take  to be the speed of light (see \cite{Espinosa:2010hh,Caprini:2015zlo} for other choices), $T_*$ -- nucleation temperature, $\alpha$ -- strength of the phase transition, and $\tilde\beta$ -- the inverse of its duration. Their values  are determined by the shape of the effective potential, thus governed by the details of the particle physics at play in the early Universe. This establishes a connection between the Lagrangian parameters of a given particle physics model and the spectrum of the resulting gravitational waves.

The contribution from sound waves is given by the empirical formula \cite{Hindmarsh:2013xza,Caprini:2015zlo}
\bea\label{swavf}
&&\hspace{-1mm}h^2 \Omega_{s}(f) \approx \nn\\
&&\frac{1.9 \times 10^{-5}}{\tilde\beta}\left(\frac{100}{g_*}\right)^{\!\frac13}\!\left(\frac{\alpha\,\kappa_s}{\alpha+1}\right)^{\!2}\!\frac{(f/f_s)^3\ \Upsilon}{\big[1+0.75\, (f/f_s)^2\big]^{\frac72}}   \ \ \  \ \ \ \ \ 
\eea
where 
\bea\label{new_swf}
\kappa_s &=& \frac{\alpha}{0.73+0.083\sqrt\alpha + \alpha} \ ,\nn\\
f_s &=& (1.9 \times 10^{-9} \ {\rm Hz} ) \left(\frac{T_*}{10 \ {\rm MeV}}\right)\left(\frac{g_*}{100}\right)^\frac16  \tilde\beta \ ,\nn\\
\Upsilon &=& 1- \frac1{\sqrt{1+\frac{8\pi^{1/3}}{\tilde{\beta}}\sqrt{\frac{\alpha+1}{3\alpha  \kappa_s}}}}
\eea
are, respectively,   the fraction of the latent heat transformed into the bulk motion of the plasma \cite{Espinosa:2010hh}, the peak frequency for $h^2\Omega_s$, and the suppression factor \cite{Ellis:2020awk,Guo:2020grp}.

The contribution to the gravitational wave spectrum from bubble wall collisions is  \cite{Kosowsky:1991ua,Huber:2008hg,Caprini:2015zlo}
\bea\label{col}
&&\hspace{-1mm}h^2 \Omega_{c}(f) \approx \nn\\
&&\frac{4.9\times 10^{-6}}{\tilde\beta^2}\left(\frac{\alpha\,\kappa_c}{\alpha+1}\right)^2\left(\frac{100}{g_*}\right)^{\!\frac13}
\frac{(f/f_c)^{2.8}}{1+2.8\, (f/f_c)^{3.8}} \ , \ \ \ \ \ \ 
\eea
where
\bea
\kappa_c &=& \frac{\frac{4}{27}\sqrt{\frac{3}{2}\alpha} + 0.72\,\alpha }{1+0.72\,\alpha}\ , \nn\\
f_c &=& (3.7\times10^{-10} \ {\rm Hz} ) \left(\frac{T_*}{10 \ {\rm MeV}}\right)\left(\frac{g_{*}}{100}\right)^\frac16\tilde\beta  \ \ \ \ \ \  
\eea
 are the fraction of the latent heat deposited into the bubble front \cite{Kamionkowski:1993fg} and the peak frequency for $h^2\Omega_c$, respectively.

Lastly, the contribution from  magnetohydrodynamic turbulence is given by  \cite{Caprini:2006jb,Caprini:2009yp},
\bea\label{turb}
h^2 \Omega_{t}(f) &\approx&  \frac{3.4\times 10^{-4}}{\tilde\beta}\left(\frac{\alpha\,\epsilon \, \kappa_s}{\alpha+1}\right)^{\frac32}  \nn\\
&\times& \left(\frac{100}{g_*}\right)^{\!\frac13} \frac{({f}/{f_t})^{3}}{\big(1+{8\pi \,f}/{h_*}\big)\big(1+{f}/{f_t}\big)^{{11}/{3}}} \ , \ \ \ \ \ \ 
\eea
in which $\epsilon$ is  the turbulence suppression  parameter that we set to  $\epsilon=0.05$ following \cite{Caprini:2015zlo}, and where
\bea
f_t &=& (2.7 \times 10^{-9}\ {\rm Hz} )\left(\frac{T_*}{10 \ {\rm MeV}}\right) \left(\frac{g_*}{100}\right)^\frac16{\tilde\beta} \ ,\nn\\
h_* &=& (1.7\times 10^{-9} \ {\rm Hz})\left(\frac{T_*}{10 \ {\rm MeV}}\right) \left(\frac{g_*}{100}\right)^\frac16\ .
\eea 
are the peak frequency and the inverse Hubble time at gravitational wave production
 redshifted to today \cite{Caprini:2015zlo}, respectively.

The combined  stochastic gravitational wave spectrum from phase transitions is obtained by adding the three contributions,
\bea\label{alll}
h^2 \Omega_{\rm GW}(f) = h^2 \Omega_{s}(f) + h^2 \Omega_{c}(f)+ h^2 \Omega_{t}(f) \ . \ \ \ \ 
\eea

\begin{figure}[t!]
\includegraphics[trim={1.7cm 0.3cm 1cm 0cm},clip,width=9.4cm]{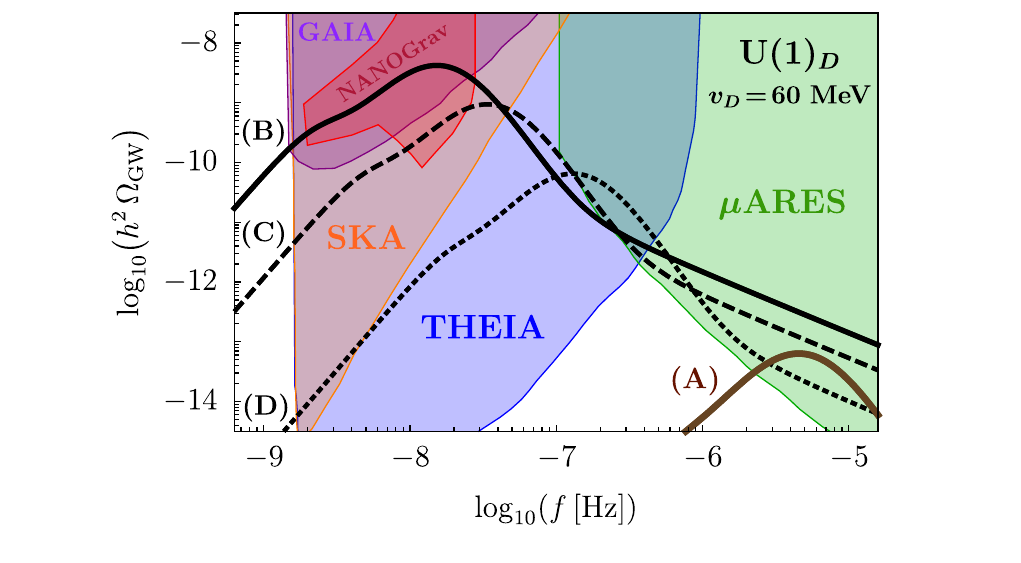} \vspace{-7mm}
\caption{Gravitational wave signatures of  a first order phase transition in the dark ${\rm U}(1)_D$ model  \cite{Cline:2018ami} assuming $v_D = 60 \ \rm MeV$, plotted for the Lagrangian parameters $(g_D, \lambda)$ specified in Table \ref{tabl}.  The shaded regions correspond to the reach of future gravitational wave and astrometry experiments: SKA \cite{Weltman:2018zrl} (orange), $\mu$ARES \cite{Sesana:2019vho} (green), THEIA \cite{Garcia-Bellido:2021zgu} (blue), GAIA \cite{Gaia1,Moore:2017ity} (purple), along with the 15-year NANOGrav signal region \cite{NANOGrav:2023hvm} (red).}\label{fig:3}
\end{figure}

Upon performing the procedure outlined in Section \ref{pt} for various values of the gauge coupling $g_D$ and the quartic coupling $\lambda$ within the  dark ${\rm U}(1)_D$ model, keeping the vacuum expectation value fixed at $v_D= 60 \ \rm MeV$ and using the empirical expressions in Eqs.\,(\ref{swavf})-(\ref{alll}), we arrived at the  representative gravitational wave signals shown in Figure \ref{fig:3}. The illustrated curves correspond to the four choices of the Lagrangian parameters: brown solid line $(\rm A)$: $(g_D,\lambda) = (0.24, 0.0001)$, black solid line $(\rm B)$: $(g_D,\lambda) = (0.8, 0.008)$, long-dashed line $(\rm C)$: $(g_D,\lambda) = (0.9, 0.012)$, and short-dashed line $(\rm D)$: $(g_D,\lambda) = (1.0, 0.018)$. 
Overplotted are the anticipated sensitivities of future experiments: pulsar timing array  SKA \cite{Weltman:2018zrl}, space-based interferometer $\mu$ARES \cite{Sesana:2019vho}, and the astrometry proposals THEIA \cite{Garcia-Bellido:2021zgu} and GAIA \cite{Gaia1,Moore:2017ity}. 
In addition, we also included the signal region suggested by the NANOGrav 15-year data  \cite{NANOGrav:2023hvm}. 

As noted earlier, in our analysis we consider the entire parameter space of the  ${\rm U}(1)_D$ model, not just the region for which a neutron dark decay channel is kinematically available. Out of the four signals plotted in Figure \ref{fig:3}, only the curve  $(\rm A)$, which  is equivalent to the benchmark point $(\rm A)$ given by Eq.\,(\ref{benchaaa}), allows for the neutron dark decay channel $n\to \chi\,\phi$ to exist with a branching fraction of $1\%$.

The central peak of each of the signals is determined predominantly by the sound  wave contribution. The small bump visible on the left side of each spectrum is due to the bubble collision contribution, whereas the change in slope on the right side of the spectra is due to the magnetohydrodynamic turbulence contribution. We note that the sound wave component would completely overwhelm the other contributions  to the spectrum if 
 not for the significant suppression coming from $\Upsilon$  in Eq.\,(\ref{swavf}), reducing the signal by a factor 
of $\mathcal{O}(10 - 100)$. Our plots take into account this large suppression, without which the signals would be much stronger.

\begin{table}[t!] 
\begin{center}
\begingroup
\setlength{\tabcolsep}{6pt} 
\renewcommand{\arraystretch}{1.5} 
\begin{tabular}{ |c|c|c |c |c|| c| c|c|} 
\hline
\!\!Signal\!\!&\multicolumn{4}{|c||}{${\rm U}(1)_D$ model parameters}  & \multicolumn{3}{|c|}{\!Transition parameters\!\!} \\ 
\hline
\hline
  &  \  {\raisebox{-1ex}{$g_D$}}  \  &\raisebox{-1ex}{$\lambda$}     & {\raisebox{0.5ex}{$m_\phi$}}  &   {\raisebox{0.5ex}{$m_{A'}$}}   &  {\raisebox{-1ex}{$\alpha$}}  & {\raisebox{-1ex}{$\tilde\beta$   }}& {\raisebox{0.5ex}{$T_*$}} \\ [-8pt]
    &  \   \  & \ \   & $\!\![\rm MeV]\!\!$  &  $\!\![\rm MeV]\!\!$  & \ \ &  \ \  & $\![\rm MeV]\!$ \\ \hline
    \hline
$ (\rm A) $&    $0.24$  &    $0.0001$  & $0.85$  &   $10$  & 0.7 & \!\! 6600 \!\! & $3.6 $ \\[1pt]
\hline
$ (\rm B) $&    $0.8$  &    $0.008$  & $7.6$  &   $34$  &  4.3 & \ 20 \ & $4.0 $ \\[1pt]
\hline
$(\rm C)$ &       $0.9$ &   $0.012$  & $9.3$  &   $38$  &     1.0    &  \ 30 \ & $5.9 $ \\[1pt]
\hline
$(\rm D)$ &     $1.0$ &  $0.018$   & $11.4$  &  $ 42$  &   0.4  &  \ 80 \ & $8.5 $  \\[1pt]
\hline
\end{tabular}
\endgroup
\end{center}
\vspace{-3mm}
\caption{The correspondence between the Lagrangian parameters $(g_D, \lambda)$ of the ${\rm U}(1)_D$  model  ($v_D= 60 \ \rm MeV$) and the resulting phase transition parameters $(\alpha, \tilde\beta, T_*)$ for the signals plotted in Figure \ref{fig:3}.}
\label{tabl}
\end{table}

\begin{figure}[t!]
\includegraphics[trim={1.2cm 0.4cm 1cm 0cm},clip,width=8.5cm]{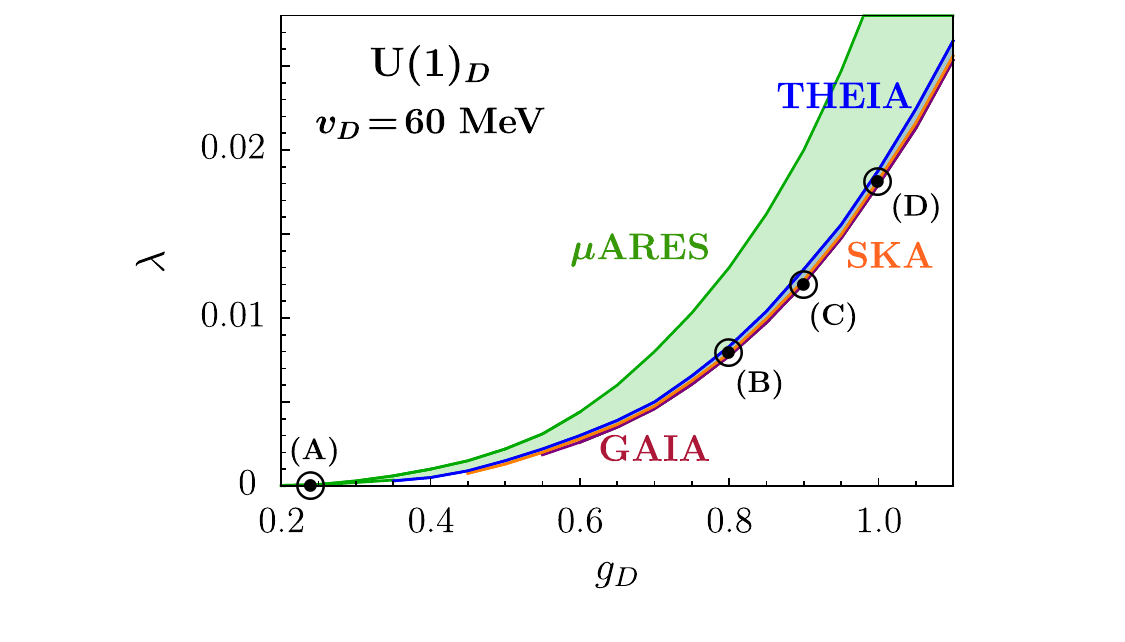} \vspace{-4mm}
\caption{The parameter space $(g_D,\lambda)$  for which the gravitational wave signal from a first order phase transition in the dark ${\rm U}(1)_D$ model is discoverable with a signal-to-noise ratio of at least five after the mission lifetime is completed for each of the experiments: GAIA (purple), THEIA (blue), SKA (orange), and $\mu$ARES (green).  The circled dots  correspond to the signal curves  in Figure \ref{fig:3}.}\label{fig:44}
\end{figure}

Table \ref{tabl} specifies the first order phase transition parameters:  strength $\alpha$, inverse of its duration $\tilde\beta$, and nucleation temperature $T_*$, corresponding to the 
Lagrangian parameters $(g_D, \lambda)$ selected for the four signal curves. It also provides the masses of the scalar $\phi$ and dark photon $A'$. The  peak frequency of the signal depends on the duration of the phase transition and the nucleation temperature, as described by Eq.\,(\ref{new_swf}); for faster phase transitions (larger $\tilde\beta$) and higher nucleation temperatures the peak shifts to higher frequencies. 
Furthermore, the height of the peak is determined by  the strength of the phase transition and its duration, resulting in a stronger signal  for larger values of $\alpha$ and for longer phase transitions, i.e.,  characterized by a smaller $\tilde\beta$.

A comment regarding the $\mathcal{O}(1-10)$ MeV   nucleation temperatures  is in place. In general, low values of $T_*$ are tightly  constrained by Big Bang nucleosynthesis and measurements of the cosmic microwave background radiation. A detailed analysis of the resulting bounds on the allowed nucleation temperatures for first order phase transitions was carried out in \cite{Bai:2021ibt,Deng:2023twb}. Depending on the strength of the transition $\alpha$, the lower limit on the nucleation temperature varies in the range $\mathcal{O}(1-2 \,\rm MeV)$. As seen from Table \ref{tabl}, all of the points we selected satisfy those bounds.

To determine how much of the parameter space of the ${\rm U}(1)_D$ model can be probed in gravitational wave and astrometry experiments, we performed a scan over $(g_D, \lambda)$. Figure \ref{fig:44} shows the regions of parameter space for the gauge coupling $g_D$ versus the quartic coupling $\lambda$ which give rise to signals detectable with a signal-to-noise ratio greater than five during the mission lifetime of  GAIA (purple), THEIA (blue), SKA (orange), and $\mu$ARES (green). 
For a given value of $g_D$, 
the larger the value of the quartic coupling $\lambda$, the weaker the signal and the higher the nucleation temperature.  Thus, for each $g_D$ the lower bound on $\lambda$ corresponds to the lowest possible nucleation temperature for that value of $g_D$, and below this bound  the value of $\mathcal{S}(T)/T$ is too large for Eq.\,(\ref{nucl_temp}) to be satisfied. The upper bound on $\lambda$ arises from the limited sensitivity of an individual experiment.  We note that all points in Figure \ref{fig:44} fulfill the Big Bang nucleosynthesis and cosmic microwave background constraints discussed in \cite{Bai:2021ibt,Deng:2023twb}. Lowering the value of $\lambda$ by too much for the benchmark point $(\rm A)$ (in order to strengthen the signal), although corresponding simply to choosing a smaller  mass of $\phi$, would lead to a lower nucleation temperature inconsistent with the discussed cosmological bounds. Therefore, model ${\rm U}(1)_D$ with the neutron dark decay channel $n\to \chi\,\phi$ open can only be probed by the $\mu$ARES experiment.

Quite remarkably, having chosen the vacuum expectation value $v_D = 60 \ \rm MeV$ that saturates the neutron star bound in Eq.\,(\ref{nns}), there exists a range of parameters  for which the signal lies in the region of interest for the NANOGrav 15-year data result, and which, at the same time, can be searched for in all other experiments considered here: GAIA, SKA, THEIA, and  $\mu$ARES. A detailed analysis of how well our signal fits the NANOGrav region is beyond the scope of this paper, especially since that case does not allow for the $n\to \chi\,\phi$ decay channel to exist. We refer the reader to a  thorough  model-independent analysis in \cite{NANOGrav:2023hvm}.
\break
\break

\begin{figure}[t!]
\includegraphics[trim={1.7cm 0.3cm 1cm 0cm},clip,width=9.4cm]{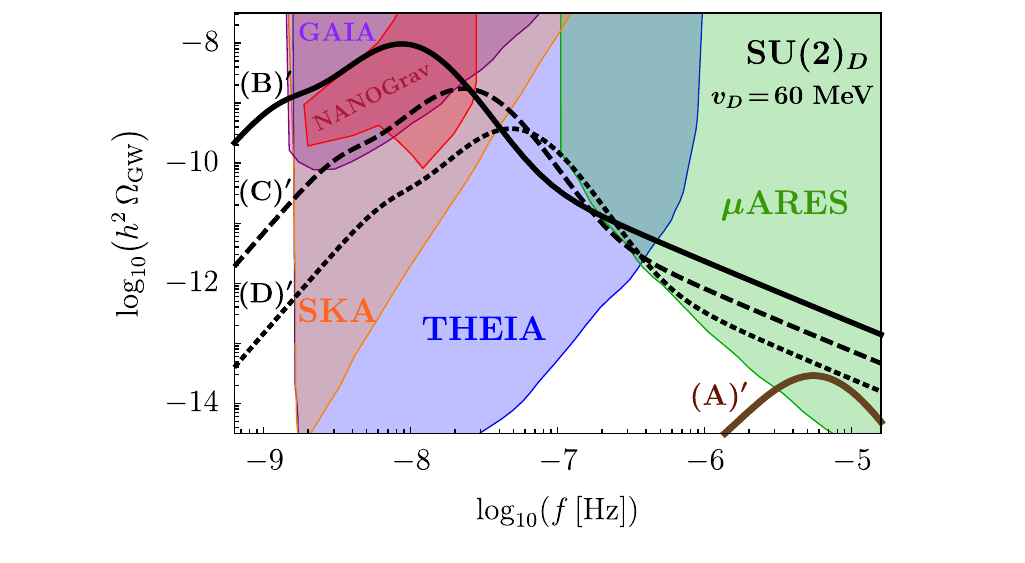} \vspace{-7mm}
\caption{Gravitational wave signals arising from a first order phase transition in the dark ${\rm SU}(2)_D$ model \cite{Elahi:2020urr} assuming $v_D = 60 \ \rm MeV$, plotted for the Lagrangian parameters $(g_D, \lambda)$ specified in Table \ref{tabl2}.  The shaded regions correspond to the predicted sensitivity of future  experiments, as explained in the caption of Figure \ref{fig:3}.}\label{fig:5}
\end{figure}

\begin{table}[t!] 
\begin{center}
\begingroup
\setlength{\tabcolsep}{6pt} 
\renewcommand{\arraystretch}{1.5} 
\begin{tabular}{ |c|c|c |c |c|| c| c|c|} 
\hline
\!\!Signal\!\!&\multicolumn{4}{|c||}{${\rm SU}(2)_D$ model parameters}  & \multicolumn{3}{|c|}{\!Transition parameters\!\!} \\ 
\hline
\hline
  &  \  {\raisebox{-1ex}{$g_D$}}  \  &\raisebox{-1ex}{$\lambda$}     & {\raisebox{0.5ex}{$m_\phi$}}  &   {\raisebox{0.5ex}{$m_{A'}$}}   &  {\raisebox{-1ex}{$\alpha$}}  & {\raisebox{-1ex}{$\tilde\beta$   }}& {\raisebox{0.5ex}{$T_*$}} \\ [-8pt]
    &  \   \  & \ \   & $\!\![\rm MeV]\!\!$  &  $\!\![\rm MeV]\!\!$  & \ \ &  \ \  & $\![\rm MeV]\!$ \\ \hline
    \hline
$ (\rm A)' $&    $0.19$  &    $0.0001$  & $0.85$  &   $8$  & 0.5 & 8700 & $3.3 $ \\[1pt]
\hline
$ (\rm B)' $&    $0.8$  &    $0.022$  & $12.6$  &   $34$  &  1.7 & \ 10 \ & $4.9 $ \\[1pt]
\hline
$(\rm C)'$ &       $0.9$ &   $0.034$  & $15.6$  &   $38$  &     0.8    &  \ 20 \ & $6.7 $ \\[1pt]
\hline
$(\rm D)'$ &     $1.0$ &  $0.050$   & $19.0$  &  $ 42$  &   0.4  &  \ 30 \ & $8.4 $  \\[1pt]
\hline
\end{tabular}
\endgroup
\end{center}
\vspace{-3mm}
\caption{Translation between the fundamental parameters $(g_D, \lambda)$ of the ${\rm SU}(2)_D$  model ($v_D= 60 \ \rm MeV$)  and the resulting phase transition parameters $(\alpha, \tilde\beta, T_*)$ for the signals plotted in Figure \ref{fig:5}.\\}
\label{tabl2}
\end{table}

\begin{figure}[t!]
\includegraphics[trim={1.4cm 0.4cm 1cm 0cm},clip,width=8.5cm]{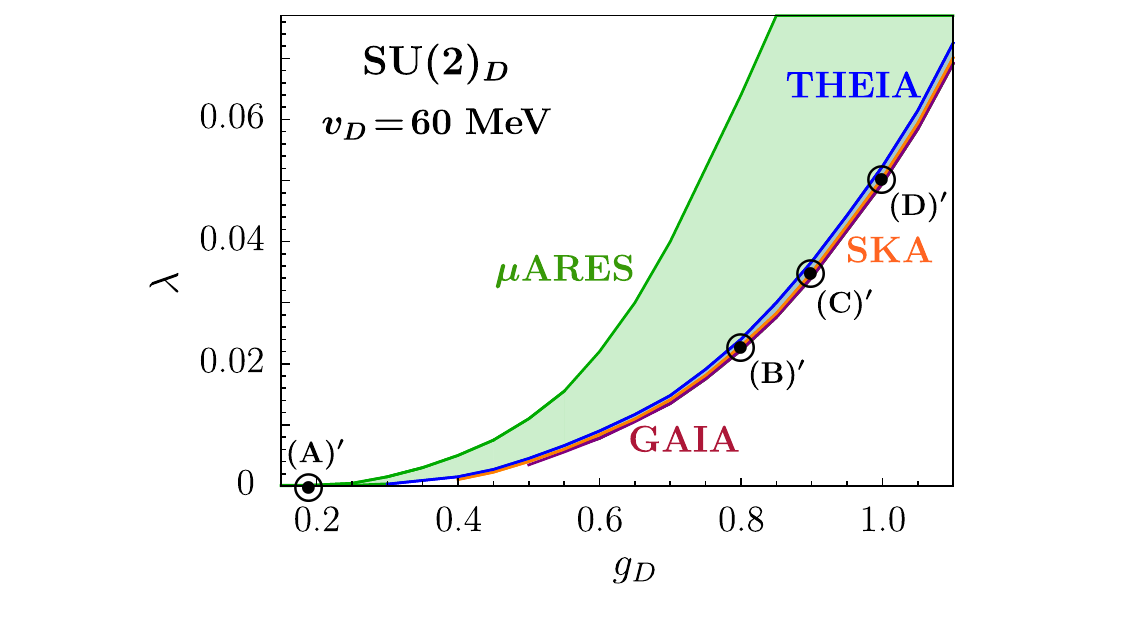} \vspace{-3mm}
\caption{Parameter space $(g_D,\lambda)$  for which the gravitational wave signal from a first order phase transition in the dark ${\rm SU}(2)_D$ model can be observed in future gravitational wave and astrometry experiments, as explained in the caption of Figure \ref{fig:44}. The circled dots  correspond to the signal curves shown in Figure \ref{fig:5}.}\label{fig:66}
\end{figure}

Performing the same analysis as  above for the dark ${\rm SU}(2)_D$ model, again keeping the vacuum expectation value fixed at $v_D = 60 \ \rm MeV$, we obtain the gravitational wave signals  shown in Figure \ref{fig:5}. For an easier comparison with the previous case, three signal lines were chosen to correspond to the same values of the gauge coupling $g_D$ as those in Figure \ref{fig:3}. The four signal curves in Figure \ref{fig:5} correspond to the following parameter choices:   brown solid line $(\rm A)'$: $(g_D,\lambda) = (0.19, 0.0001)$, black solid line $(\rm B)'$:    $(g_D,\lambda) = (0.8, 0.022)$, long-dashed line $(\rm C)'$: $(g_D,\lambda) = (0.9, 0.034)$, and short-dashed line $(\rm D)'$: $(g_D,\lambda) = (1.0, 0.050)$. 
Overplotted are  the predicted sensitivities of future gravitational wave and astrometry experiments.
Out of the four signals, only $(\rm A)'$  corresponds to the scenario allowing for the decay channel $n\to \chi\,\phi$ to exist, and is equivalent to the benchmark point $(\rm A)'$ given by Eq.\,(\ref{20}); it can be searched for only with $\mu$ARES.

Similarly as before, Table \ref{tabl2} specifies the first order phase transition parameters $(\alpha, \tilde\beta,T_*)$ which correspond to the ${\rm SU}(2)_D$ model  parameters $(g_D, \lambda)$  for the four  signal curves. The nucleation temperatures in this case are higher than for the ${\rm U}(1)_D$ model, and the discussed earlier lower bound on the nucleation temperature of $\mathcal{O}(1-2 \,\rm MeV)$ arising  from cosmological measurements is again satisfied.

Upon performing a scan over the $(g_D, \lambda)$ parameters of the model, we arrive at the results presented in  Figure \ref{fig:66}, which shows the regions of parameter space for the gauge coupling $g_D$ versus the quartic coupling $\lambda$ giving rise to signals detectable with a signal-to-noise ratio greater than five during the mission lifetime of  GAIA, THEIA, SKA, and $\mu$ARES. It is evident that in this case higher values of the quartic coupling $\lambda$ are required for the signal to be detectable. This is due to the fact that there are more degrees of freedom in the ${\rm SU}(2)_D$ model compared to the ${\rm U}(1)_D$ model, which strengthens the one-loop contribution to the effective potential, requiring a larger $\lambda$ to balance this change.

\section{Conclusions}\label{sum}

Recently, with the first successful direct detection  of gravitational waves, particle physics acquired a new and extremely  powerful tool to probe new physics models. Indeed, many extensions of the Standard Model are based on extra gauge symmetries at higher energy scales, which, when spontaneously broken, may trigger a first order phase transition leading to the emission of a stochastic gravitational wave background in the early Universe reaching us today. The shape of its spectrum is determined by the particle content of  the model, thus providing insight into the high energy structure of the theory. The literature on the subject is vast and predominantly  models with symmetry breaking scales beyond $\mathcal{O}(100 \ \rm GeV)$ have been  considered in this respect.

In this paper, we demonstrated that not only high energy particle physics, but also low-energy physics, such as nuclear physics, might benefit from the progress on the gravitational wave front. We focused on models with a symmetry breaking  scale $\mathcal{O}(60 \ \rm MeV)$   constructed to explain  the neutron lifetime anomaly, which is a puzzling discrepancy between two different types of experiments measuring the neutron lifetime. Those models, in certain regions of their parameter space, allow for  the existence of a neutron dark decay channel with dark particles in the final state, some being good dark matter candidates. We demonstrated that for a wide range of parameters those models lead to a first order phase transition in the early Universe resulting in signatures which may be detectable in future experiments searching for gravitational waves at low frequencies.

In particular, we found that for a range of parameter values in the case of a model with a dark ${\rm U}(1)_D$ gauge group and a model with a dark ${\rm SU}(2)_D$ gauge group, the stochastic gravitational wave background  arising from a first order phase transition in the early Universe falls within the reach of the future space-based  gravitational wave detector $\mu$ARES, the pulsar timing array experiment SKA, as well as the planned astrometry experiments such as GAIA and THEIA.  Since parts of the  predicted gravitational wave signals lie in regions of overlapping sensitivities of various detectors, this  offers an opportunity of cross-checking the results, encouraging stronger collaboration between the gravitational wave and astrometry communities.
Moreover, since the neutron dark decay proposal is currently being experimentally probed in various low-energy experiments, an opportunity of overlap exists also with the nuclear physics community.

In our analysis we showed that, for both models considered, there exists a range of parameter values with the signal localized in the region corresponding to the lower portion of the recent NANOGrav 15-year data set region. It would be interesting to investigate, e.g., if an additional contribution from cosmic strings would produce a better fit  to the upper part of the NANOGrav signal, and, combined with the  phase transition signal, could provide an explanation of the NANOGrav data. Such a cosmic string gravitational wave signal can naturally arise  from the breaking of the gauged lepton number ${\rm U}(1)_\ell$ symmetry at high energies, which itself could lead to the seesaw mechanism for the neutrinos, and  constitute  a natural extension of the models considered in this paper.

It goes without saying that a discovery of any primordial gravitational wave background would lead to a much needed breakthrough in particle physics.  What makes the models considered in this paper special is that they can be probed by many  different experiments, leading to multiple synergies between physics disciplines, which seems to be the most promising way forward  in our quest to discover the truth about the dark side of the Universe.

\subsection*{Acknowledgments}

This research was supported by the National Science Foundation under Grant No. PHY-2213144.

\bibliography{bibliography}

\end{document}